# The effect of collinearity and sample size on linear regression results: a simulation study

Stephanie C. C. van der Lubbe[1,2], Jose M. Valderas[1,2,3,4], Evangelos Kontopantelis[1,5]*

[1] Division of Family Medicine, Yong Loo Lin School of Medicine, National University of Singapore, Singapore
[2] National University Polyclinics, National University Health System, Singapore, Singapore
[3] Department of Family Medicine, National University Health System, Singapore, Singapore
[4] Centre for Research in Health Systems Performance (CRiHSP), National University of Singapore, Singapore
[5] Division of Informatics, Imaging and Data Sciences, Faculty of Biology, Medicine and Health, The University of Manchester, UK

* Corresponding author
Email: e.kontopantelis@manchester.ac.uk

# Abstract

**Background:** Multicollinearity inflates the variance of OLS regression coefficients, widening confidence intervals and reducing the reliability of statistical inference. Applied research commonly relies on fixed variance inflation factor (VIF) cut-offs (e.g., 4 or 10) to diagnose "problematic" collinearity and to justify combining or excluding covariates. This practice is often applied uniformly across studies with vastly different sample sizes, however, the practical consequences of collinearity are inherently finite-sample phenomena. The key question for applied work is when variance inflation becomes large enough to meaningfully compromise estimation precision, coverage, and inferential conclusions, particularly in the presence of model misspecification. We quantify how collinearity and sample size jointly shape linear regression performance across a wide spectrum of conditions, and we provide practical guidance on interpreting VIFs in context. We evaluate outcomes central to inference and decision-making.

**Methods:** We conducted a simulation study varying sample size from 100 to 100,000 and collinearity from VIF = 1 (none) to VIF = 50 (extreme). For each scenario, we generated 1,000 datasets and fitted OLS models to estimate a prespecified "main" coefficient. We assessed (i) coverage (proportion of 95% confidence intervals containing the true value), (ii) mean absolute error and bias, (iii) traditional power (proportion of 95% confidence intervals excluding 0), and (iv) precision assurance (the probability that the 95% confidence interval lies entirely within a practically meaningful margin around the true effect). We additionally evaluated a misspecified (biased) scenario by omitting a relevant predictor, to quantify how collinearity amplifies bias.

**Results:** Under correct specification, collinearity did not materially affect nominal coverage and did not introduce systematic bias, but it substantially reduced precision in small samples: at $N \approx 100$, even mild collinearity (VIF < 2) inflated errors and markedly reduced both power metrics, whereas at $N \geq 50,000$ estimates remained robust even under extreme collinearity (VIF = 50). Under misspecification, collinearity strongly amplified bias, producing larger errors, reduced coverage, and dramatic losses in both precision assurance and traditional power, even at low VIF values.

**Conclusion:** Fixed VIF thresholds should not be applied mechanically. Collinearity must be interpreted in relation to sample size and potential sources of bias: in small studies, modest VIFs can be consequential, while in large studies extreme VIFs may be practically negligible. Importantly, excluding predictors solely to reduce VIF can worsen inference by introducing or exacerbating omitted-variable bias. The heatmaps provided offer a practical reference for anticipating collinearity's impact across study sizes and modelling assumptions.

# Introduction

Collinearity refers to the presence of linear relationships among independent variables, such that one variable can be partially or completely explained by others.(1,2) In regression analysis, collinearity inflates the variance of estimated coefficients, increasing the likelihood of both Type I errors (false positives) and Type II errors (false negatives).(3,4) The amount of collinearity is often assessed via the Variance Inflation Factor (VIF), which quantifies how much the coefficient's variance is increased due to correlations with other predictors.

Despite longstanding concerns about simplistic rules of thumb,(5–7) many studies continue to apply fixed VIF thresholds (typically 4 or 10) to determine whether collinearity is problematic.(8–12) These cut-offs are often used indiscriminately without considering the sample size, leading to the same thresholds being applied in analyses ranging from fewer than 50 observations(13) to massive datasets with millions of datapoints(14). When the VIF threshold is exceeded, researchers typically address collinearity by merging correlated variables into a single predictor or excluding them entirely.(5) While these strategies eliminate collinearity from the regression model, rigid thresholds come with trade-offs: a cut-off set too high may leave problematic collinearity unaddressed, while one that is too restrictive may result in the unnecessary exclusion of important predictors.

A few previous studies have examined the impact of collinearity and sample size in linear regression models.(3,7,15,16). Simulation studies are essential in this context since the relationship between VIFs and performance metrics (e.g. power) cannot be expressed algebraically. However, these studies were limited to sample sizes of no more than 1,000 observations and explored collinearity using only a limited set of VIF values in their simulations. Furthermore, most studies did not investigate the effect of collinearity on power, an essential outcome for determining the reliability of regression results. Consequently, these studies do not capture how collinearity interacts with much larger datasets and how power is affected under different scenarios. These gaps are critical, particularly in the context of the increasing availability and use of large databases in epidemiology, health services research, biomedical data science, and other disciplines that rely on regression models to support evidence-based conclusions. Reliance on oversimplified VIF thresholds—especially when applied without regard to sample size—can lead to incorrect assessments of the impact of collinearity on regression outcomes, potentially resulting in biased or misleading conclusions about exposure–outcome relationships.

In this work, we present a comprehensive simulation study that systematically investigates how varying degrees of collinearity and sample size jointly influence linear regression outcomes across a wide range of conditions. We simulate datasets with sample sizes ranging from 100 to 100,000 and VIF values from 1 to 50, assessing the effects on coverage, error, bias, and both precision assurance (a probability-of-precision metric widely implemented in statistical software) and traditional hypothesis-testing power. While the qualitative effects of collinearity (such as variance inflation) are well known, existing guidance provides little insight into how these effects evolve across the finite-sample regime or at what thresholds collinearity becomes practically consequential for inference. By mapping this joint relationship across thousands of scenarios, including both correctly specified and mis-specified (omitted-variable) models, our findings deliver novel quantitative insights into the circumstances under which collinearity materially affects estimation accuracy, when its effects are negligible, and how these patterns differ between small-scale and large-scale datasets. The resulting heatmaps and performance surfaces provide practical, context-aware guidance for applied researchers across scientific disciplines, moving beyond rigid VIF cut-offs and toward empirically grounded assessments that explicitly incorporate sample size and underlying model structure.

# Methodology

To examine the effects of collinearity and sample size on coverage, error, bias, precision assurance and power, we generated datasets under various combinations of variance inflation factor (VIF) and sample sizes (N). This simulation approach allowed us to systematically vary sample size and collinearity while holding constant the effect sizes, error variance, and the distributional assumptions of predictors and error terms, thereby ensuring that observed effects on regression estimates were directly attributable to collinearity and sample size. Collinearity was adjusted via a wide range of VIF values, from 1 (no collinearity) to 50 (extremely high collinearity). Sample sizes varied from 100 to 100,000, capturing how statistical outcomes are affected across very small to very large samples, spanning a range of three orders of magnitude. For each combination of VIF and sample size (referred to as a scenario), 1,000 datasets were generated, providing a good balance between computational efficiency and stability of the estimated statistical metrics. Below is a summary of the dataset generation process; additional computational details are provided in the **Supplementary Methods**.

## Dataset generation

The datasets were generated as an $N \times X$ array of random values drawn from a standard normal distribution for $X$, where $N$ is the sample size and $X = [x_{\text{main}}, x_1, x_2, x_3, x_4, x_5]$. To introduce the desired level of collinearity among the predictors, we transformed $X$ using a Cholesky factor derived from a specified correlation matrix. For the datasets where all predictors were partially collinear, the off-diagonal elements $r$ in the correlation matrix were calculated as follows (see **Supporting Methods** for derivation):

$$r = \frac{R_i^2(p-2) + \sqrt{R_i^4(p-2)^2 + 4R_i^2(p-1)}}{2(p-1)} \qquad \text{Eq. 1}$$

where $p$ is the number of predictors, and $R_i^2$ is the coefficient of determination when one predictor is regressed on the other $p-1$ predictors. The error term $\varepsilon$ was generated with a mean of 0 and variance of $\pi^2/3$, matching the fixed variance of $\varepsilon$ in logistic regression under the standard logistic distribution assumption. Under the assumed mean-zero exogeneity and finite variance, OLS remains unbiased regardless of the specific error distribution; the logistic distribution is used here only to fix the error variance for comparability across scenarios. The outcome *y* was then computed using Equation 2:

$$y = \beta_0 + \beta_{\text{main}} x_{\text{main}} + \beta_1 x_1 + \beta_2 x_2 + \beta_3 x_3 + \beta_4 x_4 + \beta_5 x_5 + \varepsilon \qquad \text{Eq. 2}$$

where the true coefficients were randomly set as follows: $\beta_0 = 10$, $\beta_{\text{main}} = 2$, $\beta_1 = 1.3$, $\beta_2 = 1.5$, $\beta_3 = 6$, $\beta_4 = 3$, and $\beta_5 = 1$. This process was repeated 1,000 times for each scenario, generating 1,000 datasets of dimension $N \times 8$, with columns for $x_{\text{main}}, x_1, x_2, x_3, x_4, x_5, \varepsilon$, and $y$.

To verify the level of collinearity, we computed the Variance Inflation Factor (VIF):

$$\text{VIF}(x_i) = \frac{1}{1 - R_i^2} \qquad \text{Eq. 3}$$

## Statistical analysis

For each dataset, an ordinary least squares (OLS) linear regression was performed using Python's *statsmodels* package. This approach provides coefficient estimates along with standard errors, confidence intervals, and statistical significance tests.

To evaluate the accuracy of the estimates for the main coefficient $\beta_{\text{main}}$, we assessed coverage, error, bias, precision assurance and power for each regression. Coverage was defined as the proportion of simulations in which the true coefficient $\beta_{\text{main}}$ fell within the 95% confidence interval of the estimated coefficient $\hat{\beta}_{\text{main}}$. Coverage indicates whether the confidence interval captures the true value, but does not reflect how far the estimate may be (wide intervals can still lead to high coverage despite large errors).

Error was computed as the difference between the estimated and true values, $\hat{\beta}_{\text{main}} - \beta_{\text{main}}$. From this difference, we derived two metrics: the bias and mean absolute error (MAE). The bias captures any systematic over- or underestimation across simulations and hence reflects directional error. The MAE represents the average magnitude of the errors regardless of direction and was computed by taking the absolute value of each individual error before averaging.

We used two related power metrics: precision assurance and traditional power. In the primary approach, power was defined based on the precision of the estimate, considering the regression to have power if the confidence interval of $\hat{\beta}_{\text{main}}$ was entirely contained within a predefined range, $[\beta_{\text{main}} \pm c_{\text{powval}}]$. The value of $c_{\text{powval}}$ was calibrated such that, under the scenario *N*=1,000 and no collinearity (VIF=1), the power would be 80%, yielding a value of $c_{\text{powval}} = 0.189$. We call this precision-based power metric "precision assurance", which is also implemented in standard statistical software such as SAS and Stata. This metric is analogous to traditional power but evaluates estimation precision independently of the effect size of $\beta_{\text{main}}$.(17,18) This feature makes it particularly useful in simulation studies like ours, where the focus is on estimation accuracy rather than hypothesis testing.. As a secondary measure, power was also calculated using the conventional definition, where the regression was considered to have power if the confidence interval for $\hat{\beta}_{\text{main}}$ did not include 0. Unless otherwise stated, we report precision assurance as a more relevant metric for our research question.

The impact of the effect size was additionally assessed using Cohen's *d*, defined as $\beta_{\text{main}}/\sigma_\varepsilon$, where $\sigma_\varepsilon = \pi/\sqrt{3}$ represents the standard deviation of the error term $\varepsilon$. The effect size was varied by changing $\beta_{\text{main}}$ from 2 to 0 with 0.1 per step, while keeping $\sigma_\varepsilon$ constant. Unless otherwise stated, results in this paper were generated with $\beta_{\text{main}} = 2$, corresponding to an effect size of approximately $d \approx 1.10$.
All reported measures, including coverage, power, bias, and absolute error, were computed for the main coefficient $\beta_{\text{main}}$ as the mean and standard error across 1,000 simulations for each scenario.
All simulations were conducted using Python 3.11.5.

## 3. Results

### Coverage, absolute error, and bias

**Figure 1** presents the mean coverage, mean absolute error (MAE), and mean bias, as functions of increasing collinearity (measured by the VIF) between $x_{\text{main}}$ and $x_1$. The lines are color-coded to represent the increasing sample sizes, ranging from N=100 (dark red) to 100,000 (dark blue). The mean coverage remains approximately 95% across all sample sizes and levels of collinearity. These results demonstrate that even under extreme collinearity, the confidence intervals reliably capture the true coefficient, $\beta_{\text{main}}$, in around 95% of the time.

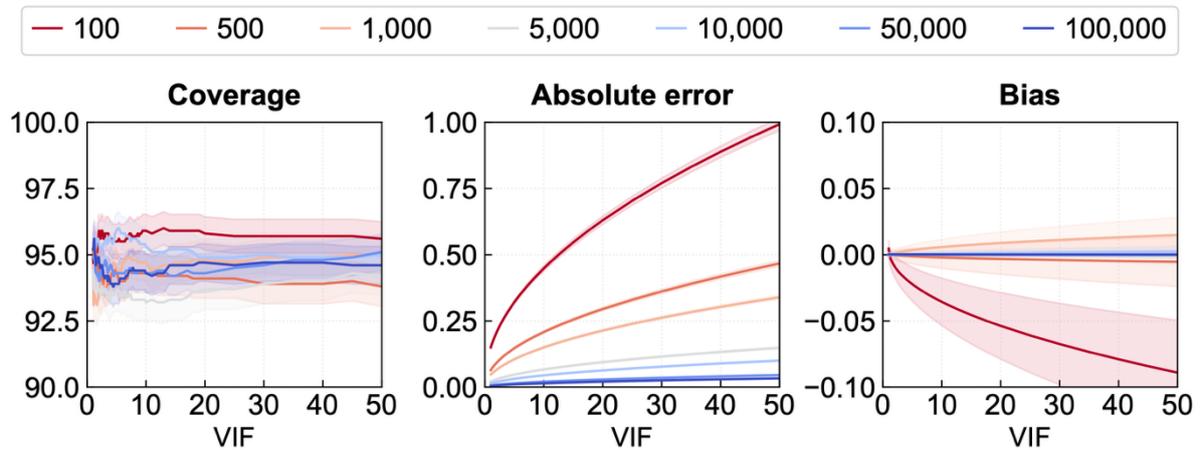

**Figure 1. Coverage, absolute error, and bias across collinearity levels.** Shown are the mean coverage (in %), mean absolute error, and mean bias for $\hat{\beta}_{\text{main}}$ as a function of collinearity, measured via the VIF, where $x_{\text{main}}$ and $x_1$ are partially collinear. Each color corresponds to a different sample size, ranging from 100 (deep red) to 100,000 (deep blue). Shaded areas represent the standard error around the mean estimates.

The MAE increases with higher VIF values, with the rate of increase depending on the sample sizes and levels of collinearity. For N=100 (the smallest sample size), collinearity has the most pronounced effect, with the MAE increasing rapidly up to a value of 0.99 at VIF=50. For larger sample sizes, the increase in MAE becomes more gradual, with N=100,000 (the largest sample size) having an MAE of only 0.03 at VIF=50.

Mean bias is near zero for all levels of collinearity for all sample sizes, except for N=100 and, to a lesser extent, N=1,000. For these smaller sample sizes, mean bias becomes more evident as collinearity increases. This bias is not due to any systematic bias in the data-generation process, but rather a consequence of finite sample variability. With smaller samples, the simulated predictors may deviate slightly from their theoretical $\mathcal{N}(0,1)$ distribution, resulting in minor shifts in the average estimated coefficients across simulations. To demonstrate that this bias stems from finite sample size, we re-calculated the bias for N=100 using a higher number of simulations per VIF. As shown in **Supplementary Figure S1**, increasing the number of simulations diminishes the bias, resulting in its near elimination. The outcome is expected, as the datasets were generated without systematic bias. Thus, collinearity alone does not introduce bias when none is present in the data. However, when bias is present, collinearity amplifies it.

# Precision assurance "power"

**Figure 2** presents the precision assurance "power" as a function of increasing collinearity (measured by the VIF) between $x_{\text{main}}$ and $x_1$, using the same color code as **Figure 1**. The dots indicate the VIF value at which precision assurance (PA) reaches 0%. As can be seen in **Figure 2**, mean PA is the most sensitive metric to collinearity and, as expected, sample size. For N=1,000, PA is benchmarked at 80% under no collinearity (VIF=1), and hence the simulations with N=100 and N=500 already have PA below 80%, without any collinearity. For N=1,000, there is 80% PA at VIF=1, which rapidly declines and reaches 0% as early as VIF=3.

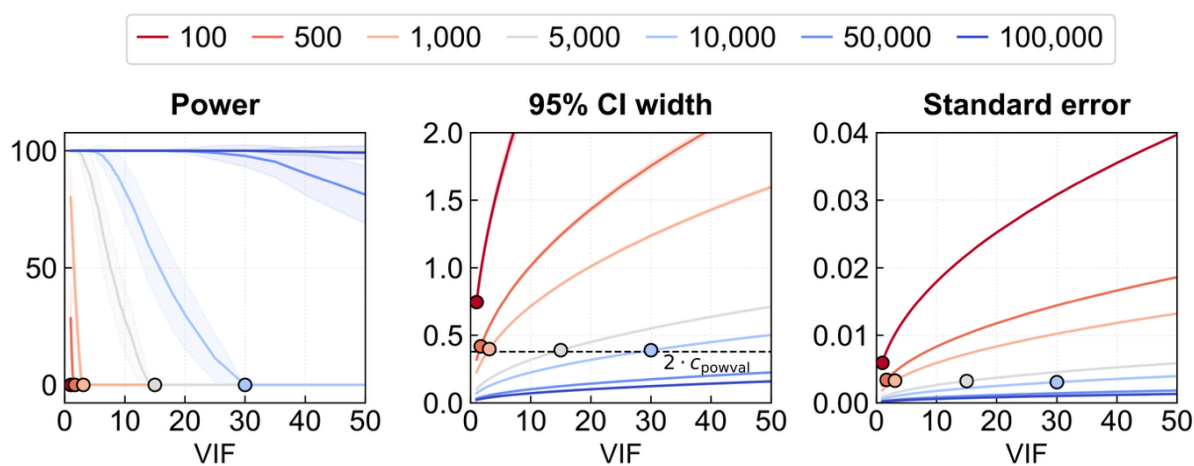

**Figure 2. Precision assurance "power", confidence interval width, and standard error across collinearity levels.** Shown are the mean precision assurance (in %), mean 95% confidence interval width, and mean standard error (SE) for $\hat{\beta}_{\text{main}}$ as a function of collinearity, measured via the VIF, where $x_{\text{main}}$ and $x_1$ are partially collinear. Each color corresponds to a different sample size, ranging from 100 (deep red) to 100,000 (deep blue). Shaded areas represent the standard error around the mean estimates.

Increasing the sample size strongly improves the robustness of PA against collinearity. For N=5,000 and N=10,000, PA drops to 0% at VIF=15 and VIF=30, respectively. For the largest sample sizes (N=50,000 and N=100,000), PA remains high across the entire VIF range, up to 50. These results underscore the critical importance of sample size in maintaining statistical power under conditions of high collinearity. The decline in PA with increasing VIF is driven by a widening of the confidence interval (**Figure 2**). As collinearity inflates the standard error of $\hat{\beta}_{\text{main}}$ (**Figure 2**), the 95% CI expands beyond $2 \cdot c_{\text{powval}}$, making it mathematically impossible to maintain PA, ultimately reducing it to 0. This mechanism is conceptually illustrated in **Figure 3**.

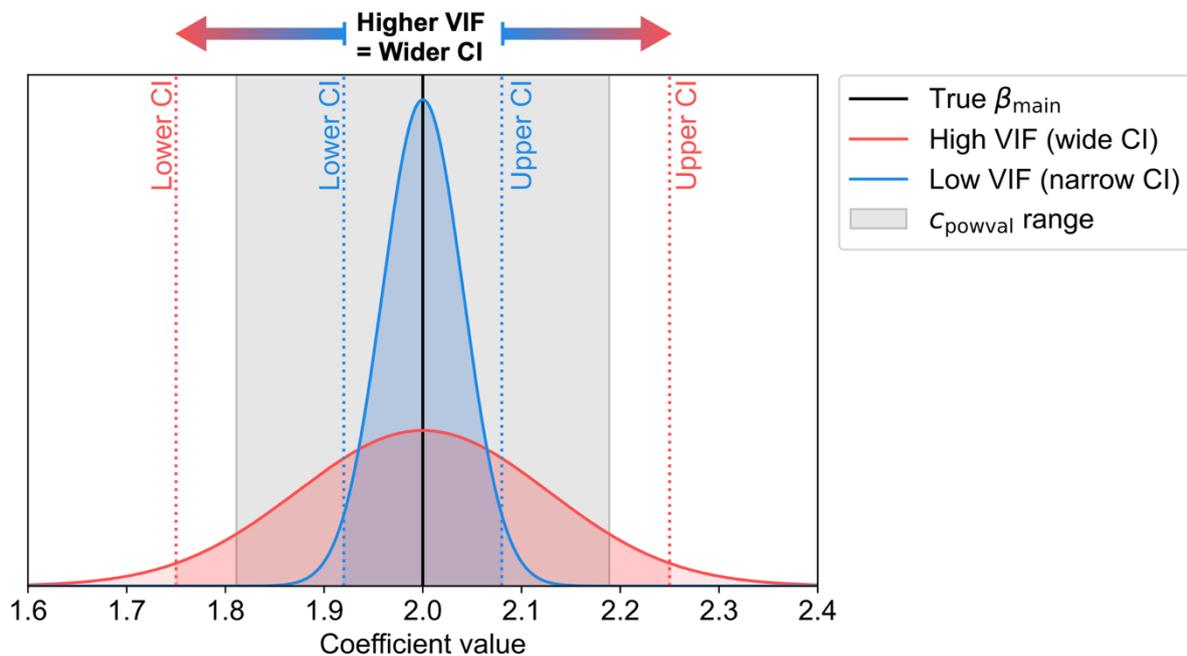

**Figure 3. Illustration of how collinearity reduces precision assurance "power".** Increasing collinearity widens the confidence interval (CI) of $\hat{\beta}_{\text{main}}$, thereby reducing the chance that the CI falls within the $c_{\text{powval}}$ range. The $c_{\text{powval}}$ range spans $\beta_{\text{main}} \pm c_{\text{powval}}$, using the same values as specified in the methodology.

## Impact of effect size

To investigate how the effect size influences the coverage, absolute error, bias, and precision assurance "power", we repeated the simulations with $\beta_{\text{main}}$ values ranging from 0 to 2 (in steps of 0.1), corresponding to Cohen's *d* values from 0 to ~1.10.

The results for coverage, MAE, and standard error remain identical across all effect sizes (**Supplementary Figure S2**). Since simulations with different $\beta_{\text{main}}$ use the same data generation processes, and the above-mentioned metrics are determined by variability in the data rather than the absolute effect size, this result is expected.

Since we defined "power" as a precision estimate (see Methodology), it remains constant across all effect sizes as well. We also examined traditional power, where the simulation is considered to have power if the 95% CI does not contain zero. As shown in **Figure 4**, traditional power decreases with increasing collinearity. However, larger effect sizes exhibit greater robustness against collinearity, as estimates further away from zero can sustain a larger standard error before their confidence intervals include zero.

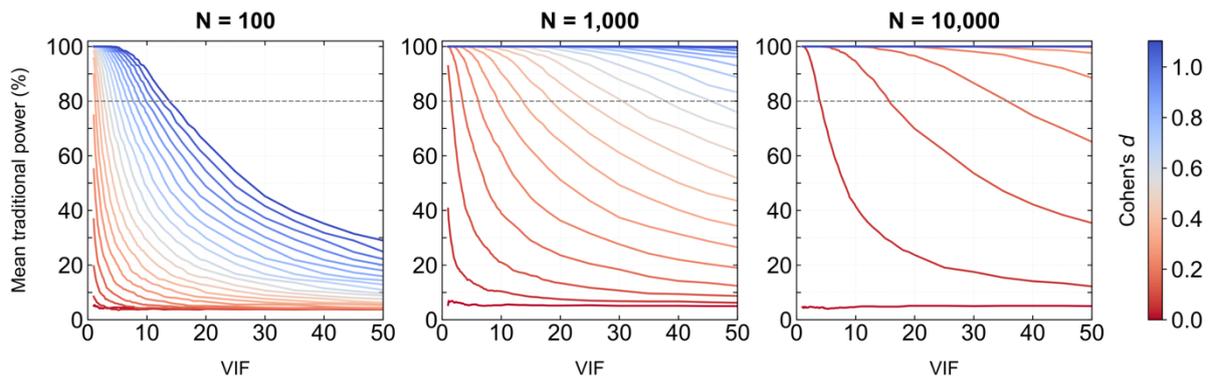

**Figure 4. Traditional power across collinearity levels and effect sizes.** Shown is the mean traditional power (in %) for $\hat{\beta}_{\text{main}}$ as a function of collinearity, measured via VIF, where $x_{\text{main}}$ and $x_1$ are partially collinear. Each panel represents a different sample size: 100 (left), 1,000 (middle), and 10,000 (right), with 1,000 simulations per scenario. Colors represent increasing values of Cohen's *d*, ranging from 0 (dark red) to ~1.10 (dark blue) in steps of 0.055.

The impact of sample size is again substantial. For N=100, the traditional power falls below 80% for all effect sizes beyond VIF=14. In contrast, for N=10,000, even small effect sizes ($d \geq 0.17$ or $\beta_{\text{main}} \geq 0.3$) maintain high power across the entire range of VIF values, even up to VIF=50 (**Figure 4**).

## Number of correlated predictors

We repeated the simulations with all predictors being partially collinear. The results show that coverage, MAE, bias, and power remain largely unchanged compared to when $x_{\text{main}}$ is correlated with only a single predictor (**Supplementary Figures S3–S4**). Thus, increasing the number of correlated predictors, while keeping the total number of predictors fixed, does not have a pronounced effect on the outcomes for $\hat{\beta}_{\text{main}}$.

All results so far were based on datasets with six independent variables (**Eq. 2**). To assess the effect of the total number of predictors, we also conducted simulations using twenty independent variables, setting $\beta_{\text{main}} = 2$ and assigning values between 0 and 9 to $[\beta_1, \ldots, \beta_{19}]$ (see **Supplementary Table S1**), with all predictors partially collinear. The results for coverage, MAE, bias, and power remained approximately unchanged compared to the simulations using six predictors (**Supplementary Figures S5–S6**). Hence, neither the total number of predictors nor the number of partially collinear predictors has a substantial effect on the outcomes for $\beta_{\text{main}}$, as opposed to sample size, which has a critical impact.

# Discussion

Contrary to widely used approaches, collinearity is not necessarily problematic. Even with extreme collinearity (VIF = 50), coverage remained unaffected, and no bias was introduced, provided that the underlying uncorrelated data were unbiased. While collinearity did inflate the variance of estimated coefficients—leading to increased absolute errors, wider confidence intervals, and reduced statistical power (traditional or precision based)—the magnitude of these effects was highly dependent on sample size.

## Strengths and limitations

A major strength of this study is its systematic, high-resolution exploration of how collinearity and sample size jointly influence linear regression inference across a wide range of realistic scenarios. While asymptotic theory implies that sampling variability decreases with $1/\sqrt{N}$ and thus attenuates the impact of a fixed VIF as $N$ grows, applied analyses operate in finite samples where the rate of this attenuation and the practical transition boundaries remain the key operational question addressed here. By evaluating multiple inferential targets (coverage, error, bias, precision and both conventional hypothesis-testing power and a precision assurance metric), our results provide an integrated view of when collinearity is practically consequential and when it is largely negligible. In addition, we explicitly examine both a correctly specified model and a mis-specified (omitted-variable) scenario, which illustrates how collinearity can interact with bias mechanisms that are common in applied research.

Several limitations should be considered when interpreting these findings. First, our primary simulations were generated under idealised conditions, assuming a correctly specified linear model with independent mean-zero errors and an error term independent of the predictors. Under these assumptions, OLS estimates are unbiased, meaning there is no systematic bias for collinearity to amplify; any small apparent bias observed in the smallest samples reflects finite-sample variability and Monte Carlo error rather than a violation of OLS unbiasedness. Second, real-world datasets frequently depart from these ideal conditions: bias can arise from measurement error, omitted variables, model misspecification (e.g., nonlinearity), heteroskedasticity, or clustering, and collinearity may exacerbate these problems by making it harder to separate predictor effects. For example, when a relevant variable is omitted, its effect may be absorbed by remaining predictors, particularly those that are highly collinear, leading to biased coefficients and degraded inference. Third, because "power" is commonly understood as hypothesis-testing power, we avoid ambiguity by referring to our primary probability-of-precision metric as precision assurance (i.e., the probability that the 95% confidence interval lies within a practically meaningful margin around the true effect), and we report conventional power alongside it to address both precision and statistical significance goals. Finally, our focus is on OLS because fixed VIF cut-offs are frequently used to justify variable exclusion in that setting; extensions to penalised regression and other modelling frameworks may yield different trade-offs and represent natural directions for future work.

## Implications

At large sample sizes (N≥50,000 in our simulations), power was retained, as measured by both relevant metrics; absolute errors remained small, and confidence intervals stayed relatively narrow across all levels of collinearity. Nevertheless, predictors with high VIF values are often excluded based on rigid threshold-based criteria, potentially omitting variables that contribute significantly to the outcome. In contrast, in small samples of only a few hundred observations, even mild collinearity (e.g., VIF=2) substantially reduced both traditional power and precision assurance. Under such conditions, meaningful predictors may be incorrectly deemed statistically not significant, increasing the risk of Type II errors. This strong dependence on sample size challenges the validity of commonly used VIF thresholds such as 5 or 10, and underscores the importance of evaluating collinearity in the context of sample size rather than relying on arbitrary cutoffs.

To provide practical guidance, we generated a heatmap illustrating the interplay between sample size, VIF, and precision assurance "power" (**Figure 5**). This visualization aids in the evaluation of whether collinearity is likely to impact power within a given study design. Since power in this context is defined as a precision estimate (see Methodology), benchmarked at 80% for N=1000 and VIF=1, the heatmap is independent of the effect size. Additional heatmaps displaying traditional power for $\beta_{\text{main}} \in \{0.5, 1, 1.5, 2\}$ are given in the **Supplementary Figures S7–S10**, offering further reference points for applied research.

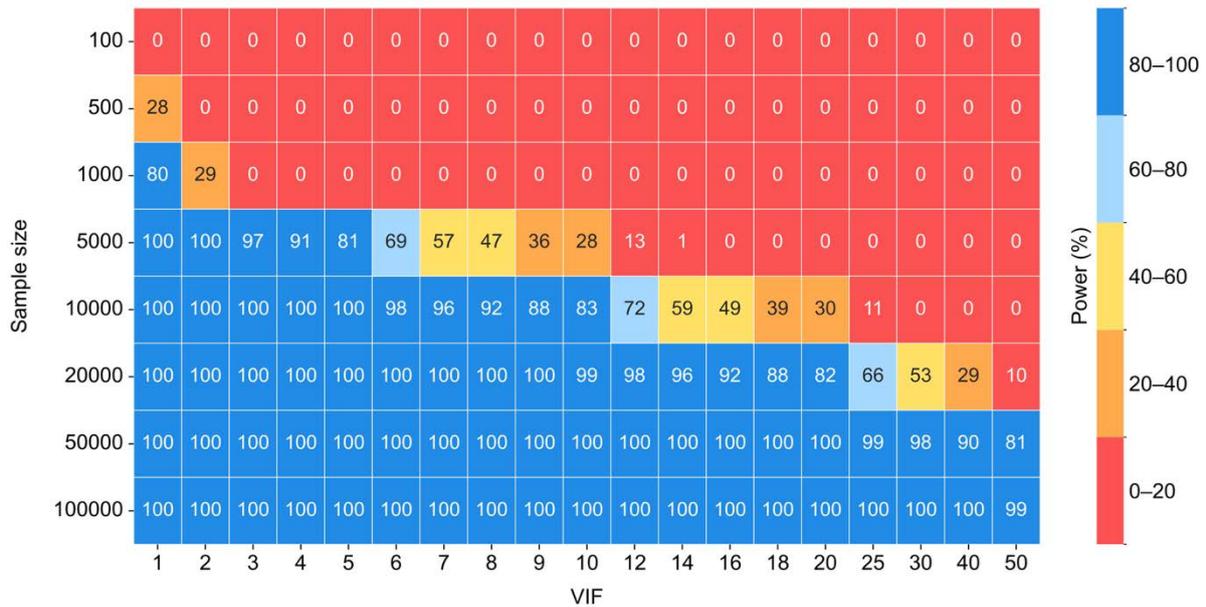

**Figure 5**. **Heatmap showing precision assurance "power" across collinearity levels and sample sizes.** Shown is the mean precision assurance "power" (in %) for $\hat{\beta}_{\text{main}}$ as a function of collinearity, measured via the VIF (x-axis), where $x_{\text{main}}$ and $x_1$ are partially collinear, and the sample size (y-axis).

The impact of collinearity becomes substantially more severe when bias is present. To illustrate this interaction, we repeated the linear regression analyses on the datasets in which $x_4$—a significant predictor with $\beta_4 = 3$—was omitted, thereby introducing bias. As shown in **Figures S11–S13**, this bias is strongly amplified by the presence of collinearity, resulting in greater estimation errors, reduced coverage, and a dramatic loss in power metrics. Power, traditional or precision-based, is particularly affected. For example, even minimal collinearity (VIF ≈ 1.1) is sufficient to reduce precision assurance "power" to zero in the biased dataset (**Figure S12**). This drop occurs because bias simultaneously shifts coefficient estimates away from the true value and widens confidence intervals, making it less likely that the 95% confidence intervals fall within the $\beta_{\text{main}} \pm c_{\text{powval}}$ range (**Figure S13**).

An additional takeaway from **Figures S11–S12** is that omitting a correlated predictor to eliminate collinearity results in substantially worse coefficient estimates than including the correlated predictor in the model. These findings, consistent with earlier work by Tay(19), reinforce the idea that while collinearity is often seen as problematic, it may be less harmful than the biases introduced by variable omission**.** Given these considerations, researchers should interpret **Figure 5** with caution, as the heatmap-based guidance was generated under ideal conditions, assuming no bias. In applied research, where collected data may be biased, collinearity may amplify these biases, leading to statistical distortions more severe than those captured by our simulations on the unbiased datasets. As such, fixed

VIF thresholds should not be applied without considering both sample size and potential sources of bias in the data.

# Conclusion

We conducted a systematic simulation study to assess how collinearity affects linear regression estimates, varying sample sizes from 100 to 100,000 and VIF values from 1 to 50. While collinearity does not affect coverage or introduce bias, it inflates the variance of estimated coefficients, leading to increased absolute errors, wider confidence intervals, and reduced statistical power (traditional or precision-based). Importantly, the magnitude of these effects depends heavily on sample size. In very small datasets (N≈100), even mild collinearity (VIF<2), which is well below commonly used thresholds, was already problematic. In contrast, for large datasets (N≥50,000), coefficient estimates remained reliable even under extreme collinearity.

These considerations lead to several practical recommendations. First, researchers should avoid mechanically applying fixed VIF thresholds, as their practical implications depend strongly on sample size. In small samples, even mild collinearity (VIF < 2) substantially reduces estimation precision, whereas in very large samples even extreme collinearity (VIF = 50) has negligible practical impact. Second, collinearity should not be interpreted in isolation: when bias is present (whether due to omitted variables, measurement error, or misspecification) collinearity can dramatically magnify its effects. Third, removing predictors solely to reduce VIF is rarely advisable; this often worsens inference by introducing or amplifying omitted-variable bias. When collinearity remains a concern, alternative modelling strategies such as penalised regression or pre-specification of variable sets may be preferable to variable exclusion. Finally, users of our heatmaps and guidance should recognise that they reflect idealised unbiased conditions; in applied settings, potential bias sources must be weighed alongside sample size and VIF.

# Acknowledgements

This work was partially funded by the DFM Research Capabilities Building Budget under the project "Technology and Compassion: Improving Patient Outcomes Through Data Analytics and Patients' Voice in Primary Care" [NUHSRO/2022/049/NUSMed/DFM]. EK was partially funded by The Manchester British Heart Foundation Centre of Research Excellence (CRE).

# 1. Supplementary Methods: Generation of datasets

**S1.1**   General procedure to generate correlated datasets

**S1.2**   Generation of correlation matrices

**S1.3**   Derivation of Equation 1 (S5 in Supporting Information)

**Figure S1.** Mean bias of $\hat{\beta}_{\text{main}}$ as a function of the collinearity, measured via VIF, with $x_{\text{main}}$ and $x_1$ being partially collinear. Each color represents a different number of simulations per scenario (i.e., each VIF/sample size combination), with a fixed sample size of 100 for each scenario. Shaded areas represent the standard error around the mean estimates.

**Figure S2.** Mean coverage (in %), mean absolute error, and mean bias for $\hat{\beta}_{\text{main}}$ as a function of the collinearity, measured via the VIF, where $x_{\text{main}}$ and $x_1$ are partially collinear, with a sample size of 1,000. Solid light coral line represents $\beta_{\text{main}} = 0$, while the dashed black line represents $\beta_{\text{main}} = 2$.

**Figure S3.** Mean coverage (in %), mean absolute error, and mean bias for $\hat{\beta}_{\text{main}}$ as a function of the collinearity, measured via the VIF, with all predictors being partially collinear. Each color corresponds to a different sample size, varying from 100 (deep red) to 100,000 (deep blue). Shaded areas represent the standard error around the mean estimates.

**Figure S4**. Mean power (in %), mean 95% confidence interval width, and mean standard error (SE) for $\hat{\beta}_{\text{main}}$ as a function of the collinearity, measured via the VIF, with all predictors being partially collinear. Each color corresponds to a different sample size, varying from 100 (deep red) to 100,000 (deep blue). Shaded areas represent the standard error around the mean estimates.

**Figure S5.** Mean coverage (in %), mean absolute error, and mean bias for $\hat{\beta}_{\text{main}}$ as a function of the collinearity, measured via the VIF. Simulations were performed with 20 independent variables, all partially collinear. Each color represents a different sample size, ranging from 100 (deep red) to 100,000 (deep blue). Shaded areas represent the standard error around the mean estimates.

**Figure S6.** Mean power (in %), mean 95% confidence interval width, and mean standard error (SE) for $\hat{\beta}_{\text{main}}$ as a function of the collinearity, measured via the VIF. Simulations were performed with 20 independent variables, all partially collinear. Each color represents a different sample size, ranging from 100 (deep red) to 100,000 (deep blue). Shaded areas represent the standard error around the mean estimates.

**Figure 7**. Mean traditional power (in %) for $\hat{\beta}_{\text{main}}$ as a function of the collinearity, measured via the VIF (x-axis), where $x_{\text{main}}$ and $x_1$ are partially collinear, and the sample size (y-axis), with $\boldsymbol{\beta}_{\text{main}} = \mathbf{2.0}$. Traditional power is defined as the proportion of simulations where the confidence interval does not include 0.

**Figure 8**. Mean traditional power (in %) for $\hat{\beta}_{\text{main}}$ as a function of the collinearity, measured via the VIF (x-axis), where $x_{\text{main}}$ and $x_1$ are partially collinear, and the sample size (y-axis), with $\boldsymbol{\beta}_{\text{main}} = \mathbf{1.5}$. Traditional power is defined as the proportion of simulations where the confidence interval does not include 0.

**Figure 9**. Mean traditional power (in %) for $\hat{\beta}_{\text{main}}$ as a function of the collinearity, measured via the VIF (x-axis), where $x_{\text{main}}$ and $x_1$ are partially collinear, and the sample size (y-axis), with $\boldsymbol{\beta}_{\text{main}} = \mathbf{1.0}$. Traditional power is defined as the proportion of simulations where the confidence interval does not include 0.

**Figure 10**. Mean traditional power (in %) for $\hat{\beta}_{\text{main}}$ as a function of the collinearity, measured via the VIF (x-axis), where $x_{\text{main}}$ and $x_1$ are partially collinear, and the sample size (y-axis), with $\boldsymbol{\beta}_{\text{main}} = \mathbf{0.5}$. Traditional power is defined as the proportion of simulations where the confidence interval does not include 0.

**Figure S11.** Mean coverage (in %), mean absolute error, mean 95% confidence interval width, and mean bias for $\hat{\beta}_{\text{main}}$ as a function of collinearity, measured via the VIF. The dataset is generated with all six predictors being partially collinear, after which one predictor ($x_4$) is omitted, introducing bias in the dataset. Each color represents a different sample size, ranging from 100 (deep red) to 100,000 (deep blue).

**Figure S12.** Mean power (in %) for $\hat{\beta}_{\text{main}}$ as a function of collinearity, measured via the VIF (X-axis), and sample size (y-axis). The dataset is generated with all six predictors being partially collinear, after which one predictor ($x_4$) is omitted, introducing bias into the dataset.

**Figure S13.** Schematic representation of why bias results in a loss of power. For an unbiased dataset, the 95% confidence interval (CI) of $\hat{\beta}_{\text{main}}$ (blue curve) falls within the $\beta_{\text{main}} \pm c_{\text{powval}}$ range, which is defined as having power. However, when bias is introduced, the estimates of $\hat{\beta}_{\text{main}}$ shift away from the true value and have wider CIs (red curves). Both effects contribute to the 95% CI of $\hat{\beta}_{\text{main}}$ falling outside the $\beta_{\text{main}} \pm c_{\text{powval}}$ range, resulting in a loss of power.

**Table S1**. True coefficients used in the simulations with 20 independent variables.

# Supplementary Information

# The effect of collinearity and sample size on linear regression results: a simulation study

Stephanie C. C. van der Lubbe[1,2], Jose M. Valderas[1,2,3,4], Evangelos Kontopantelis[1,5]

[1]    Division of Family Medicine, Yong Loo Lin School of Medicine, National University of Singapore, Singapore
[2]    National University Polyclinics, National University Health System, Singapore, Singapore
[3]    Department of Family Medicine, National University Health System, Singapore, Singapore
[4]    Centre for Research in Health Systems Performance (CRiHSP), National University of Singapore, Singapore
[5]    Division of Informatics, Imaging and Data Sciences, Faculty of Biology, Medicine and Health, University of Manchester, UK

## S1.  Supplementary Methods: Generation of datasets

*S1.1  General procedure to generate correlated datasets*

The datasets were generated using the following procedure:

Step 1:  Generation of correlation matrix
The correlation matrix was generated to yield the desired VIF level of interest for each predictor. For methodological details on how the off-diagonal values were determined, see *Section S2: Generation of correlation matrices* on the next page.

Step 2:  Calculation of the Cholesky factor from the correlation matrix
The Cholesky factor was obtained using the `linalg` module from the NumPy package (version 1.26.4) in Python (version 3.11.5), based on the correlation matrix from Step 1:

```
np.linalg.cholesky(correlation_matrix)
```

Step 3:  Generation of $N \times X$ array with random values
An array of shape $N \times X$ was generated by drawing independent values from a standard normal distribution using the `random` module from the NumPy package:

```
np.random.normal(size=(N, 6))
```

Each column corresponds to one of the predictors in $X = [x_{\text{main}}, x_1, x_2, x_3, x_4, x_5]$, and each row represents a single observation, with $N$ observations in total. We used the same set of random seeds (from 0 to 999) for each scenario, generating identical sets of 1,000 $N \times X$ arrays before introducing the desired level of collinearity. This approach ensures that any observed changes in the estimated coefficients are entirely attributable to the effect of collinearity, rather than to random differences in the underlying data.

Step 4:  Transformation using the Cholesky factor
The $N \times X$ array was then multiplied by the Cholesky factor. This operation transforms the initially independent variables into a new set of predictors with the desired correlation structure, introducing the intended collinearity.

Step 5:  Generation of error term $\varepsilon$
The error term $\varepsilon$ was drawn from a normal distribution with mean of 0 and variance of $\pi^2/3$, using the `random` module from the NumPy package:

```
np.random.normal(0, np.pi / np.sqrt(3), N)
```

Step 6: Calculation of outcome $y$
The outcome $y$ was calculated using the following linear regression equation:

$$y = \beta_0 + \beta_{\text{main}} x_{\text{main}} + \beta_1 x_1 + \beta_2 x_2 + \beta_3 x_3 + \beta_4 x_4 + \beta_5 x_5 + \varepsilon \qquad \text{Eq. S1}$$

with true coefficients set to:

$$\beta_0 = 10, \beta_{\text{main}} = 2, \beta_1 = 1.3, \beta_2 = 1.5, \beta_3 = 6, \beta_4 = 3, \text{ and } \beta_5 = 1.$$

**Step 3 to 6** were repeated 1,000 times for each scenario (i.e., each combination of sample size and VIF), resulting in 1,000 datasets of dimensions $N \times 8$, with columns representing $x_{\text{main}}, x_1, x_2, x_3, x_4, x_5, \varepsilon$, and $y$.

*S1.2   Generation of correlation matrices*

To introduce the desired level of collinearity, we generated correlation matrices corresponding to the targeted VIF values. For datasets with six variables in which only $x_{\text{main}}$ and $x_1$ are partially collinear, the 6x6 correlation matrix $\mathbf{R}_{\text{corr}}$ takes the following form:

$$\mathbf{R}_{\text{corr}} = \begin{bmatrix} 1 & r & 0 & 0 & 0 & 0 \\ r & 1 & 0 & 0 & 0 & 0 \\ 0 & 0 & 1 & 0 & 0 & 0 \\ 0 & 0 & 0 & 1 & 0 & 0 \\ 0 & 0 & 0 & 0 & 1 & 0 \\ 0 & 0 & 0 & 0 & 0 & 1 \end{bmatrix}$$

The Pearson correlation coefficient $r$ was directly derived from the definition of the Variance Inflation Factor (VIF):

$$\text{VIF}(x_i) = \frac{1}{1 - R_i^2} = \frac{1}{1 - \rho_{x_{\text{main}}, x_1}^2} = \frac{1}{1 - r^2} \qquad \text{Eq. S2}$$

where $\rho_{x_{\text{main}}, x_1}$ is the Pearson correlation coefficient between $x_{\text{main}}$ and $x_1$, and corresponds to the off-diagonal elements $r$ in the correlation matrix. Equation S2 can be rearranged to obtain $r$:

$$r = \sqrt{1 - \frac{1}{\text{VIF}(x_i)}} \qquad \text{Eq. S3}$$

For datasets in which all predictors are partially correlated, the correlation matrix is given by

$$\mathbf{R}_{\text{corr}} = \begin{bmatrix} 1 & r & r & r & r & r \\ r & 1 & r & r & r & r \\ r & r & 1 & r & r & r \\ r & r & r & 1 & r & r \\ r & r & r & r & 1 & r \\ r & r & r & r & r & 1 \end{bmatrix}$$

For this equicorrelation matrix, where all off-diagonals have the same value for $r$, we used the following analytical expression to solve for $r$:

$$r = \frac{R_i^2(p-2) + \sqrt{R_i^4(p-2)^2 + 4R_i^2(p-1)}}{2(p-1)} \qquad \text{Eq. S4}$$

where $p$ is the number of predictors, and $R_i^2$ is the coefficient of determination when one predictor is regressed on the other $p - 1$ predictors. A derivation of this equation is given in the next section (*Section S3: Derivation of Equation S4*).

The off-diagonal values of the 6x6 correlation matrices corresponding to each targeted VIF value are provided below as (VIF, $r$) pairs:

```
(1.0, 0.000), (1.1, 0.176), (1.2, 0.261), (1.3, 0.326), (1.4, 0.380),
(1.5, 0.424), (1.6, 0.462), (1.7, 0.496), (1.8, 0.525), (1.9, 0.551),
(2.0, 0.574), (2.1, 0.595), (2.2, 0.614), (2.3, 0.631), (2.4, 0.647),
(2.5, 0.661), (2.6, 0.674), (2.7, 0.687), (2.8, 0.699), (2.9, 0.709),
(3.0, 0.719), (3.1, 0.728), (3.2, 0.737), (3.3, 0.745), (3.4, 0.752),
(3.5, 0.759), (3.6, 0.766), (3.7, 0.773), (3.8, 0.779), (3.9, 0.785),
(4.0, 0.790), (4.1, 0.795), (4.2, 0.800), (4.3, 0.804), (4.4, 0.809),
(4.5, 0.814), (4.6, 0.818), (4.7, 0.821), (4.8, 0.825), (4.9, 0.829),
(5.0, 0.832), (5.2, 0.839), (5.4, 0.845), (5.6, 0.850), (5.8, 0.856),
(6.0, 0.860), (6.2, 0.865), (6.4, 0.869), (6.6, 0.873), (6.8, 0.877),
(7.0, 0.880), (7.2, 0.884), (7.4, 0.887), (7.6, 0.890), (7.8, 0.893),
(8.0, 0.895), (8.2, 0.898), (8.4, 0.900), (8.6, 0.903), (8.8, 0.905),
(9.0, 0.907), (9.2, 0.909), (9.4, 0.911), (9.6, 0.913), (9.8, 0.915),
(10, 0.916), (11, 0.924), (12, 0.931), (13, 0.936), (14, 0.941),
(15, 0.944), (16, 0.948), (17, 0.951), (18, 0.953), (19, 0.956),
(20, 0.958), (25, 0.967), (30, 0.972), (35, 0.976), (40, 0.979),
(45, 0.982), (50, 0.983)
```

### S1.3 Derivation of Equation 1 (S5 in Supporting Information)

To derive the analytical expression for $r$ in Equation S4, we begin with the definition of the squared multiple correlation coefficient $R_i^2$, which quantifies the proportion of variance explained when regressing one predictor on the remaining $p - 1$ predictors:

$$R_i^2 = \vec{r}^{\mathsf{T}} \mathbf{R}^{-1} \vec{r} \qquad \text{Eq. S5}$$

Here, $\vec{r}$ is the vector of correlations between the predictor of interest and the remaining $p - 1$ predictors, and $\mathbf{R}$ is the correlation matrix among those $p - 1$ predictors.[S1] The vector $r$ can be rewritten in terms of $r \cdot \vec{1}$, where $\vec{1}$ is an all-ones vector of length $p - 1$:

$$R_i^2 = r^2 \cdot \vec{1}^{\mathsf{T}} \mathbf{R}^{-1} \vec{1} \qquad \text{Eq. S6}$$

The inverse of the correlation matrix is given by

$$\mathbf{R}^{-1} = \frac{1}{1-r}\left(\mathbf{I}_{p-1} - \frac{r}{1+(p-2)r}\mathbf{J}_{p-1}\right) \qquad \text{Eq. S7}$$

where $\mathbf{I}_{p-1}$ is the identify matrix and $\mathbf{J}_{p-1} = \vec{1}_{p-1} \cdot 1_{p-1}^{\mathsf{T}}$ is a matrix of ones.[S2] By plugging in Equation S7 into Equation S6, we obtain:

$$R_i^2 = r^2 \cdot \vec{1}^{\mathsf{T}} \left[\frac{1}{1-r}\left(\mathbf{I}_{p-1} - \frac{r}{1+(p-2)r}\mathbf{J}_{p-1}\right)\right]\vec{1} \qquad \text{Eq. S8}$$

By applying the following algebraic simplifications:

- $\vec{1}_{p-1}^{\mathsf{T}} \mathbf{I}_{p-1} \vec{1}_{p-1} = 1_{p-1}^{\mathsf{T}} \vec{1}_{p-1} = p - 1$
- $\vec{1}_{p-1}^{\mathsf{T}} \mathbf{J}_{p-1} \vec{1}_{p-1} = \vec{1}_{p-1}^{\mathsf{T}} \vec{1}_{p-1} \cdot 1_{p-1}^{\mathsf{T}} \vec{1}_{p-1} = (p-1)^2$

Equation 8 can be rewritten to:

$$R_i^2 = \frac{r^2}{1-r}\left((p-1) - \frac{r(p-1)^2}{1+(p-2)r}\right) \qquad \text{Eq. S9}$$

which may be further simplified to:

$$R_i^2 = \frac{(p-1)r^2}{1+(p-2)r} \qquad \text{Eq. S10}$$

Rewriting Equation S10 into the quadratic form $ar^2 + br + c = 0$ and solving for $r$ using the quadratic formula yields the final analytical equation used to calculate the matrix elements $r$ in the equicorrelation matrix:

$$r = \frac{R_i^2(p-2) + \sqrt{R_i^4(p-2)^2 + 4R_i^2(p-1)}}{2(p-1)}$$

**Supplementary References**

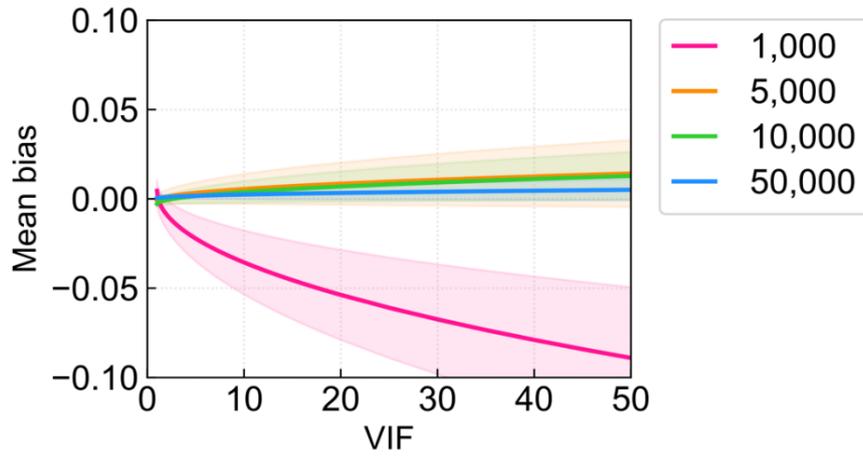

**Figure S1 |** Mean bias of $\hat{\beta}_{\text{main}}$ as a function of the collinearity, measured via VIF, with $x_{\text{main}}$ and $x_1$ being partially collinear. Each color represents a different number of simulations per scenario (i.e., each VIF/sample size combination), with a fixed sample size of 100 for each scenario. Shaded areas represent the standard error around the mean estimates.

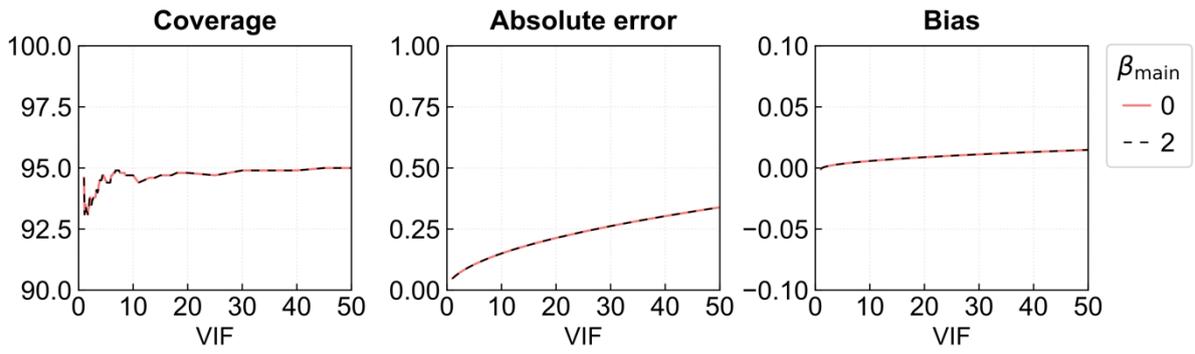

**Figure S2** | Mean coverage (in %), mean absolute error, and mean bias for $\hat{\beta}_{\text{main}}$ as a function of the collinearity, measured via the VIF, where $x_{\text{main}}$ and $x_1$ are partially collinear, with a sample size of 1,000. Solid light coral line represents $\beta_{\text{main}} = 0$, while the dashed black line represents $\beta_{\text{main}} = 2$.

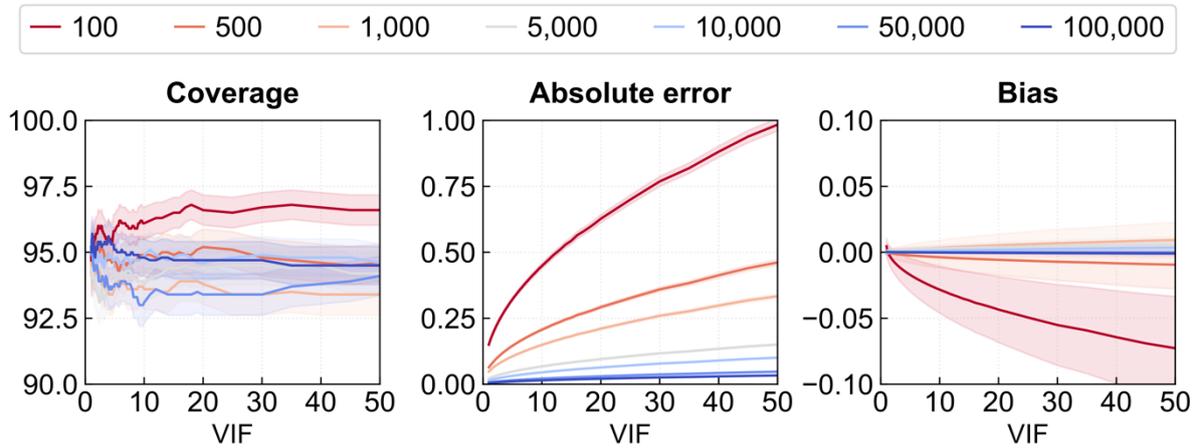

**Figure S3** | Mean coverage (in %), mean absolute error, and mean bias for $\hat{\beta}_{\text{main}}$ as a function of the collinearity, measured via the VIF, with all predictors being partially collinear. Each color corresponds to a different sample size, varying from 100 (deep red) to 100,000 (deep blue). Shaded areas represent the standard error around the mean estimates.

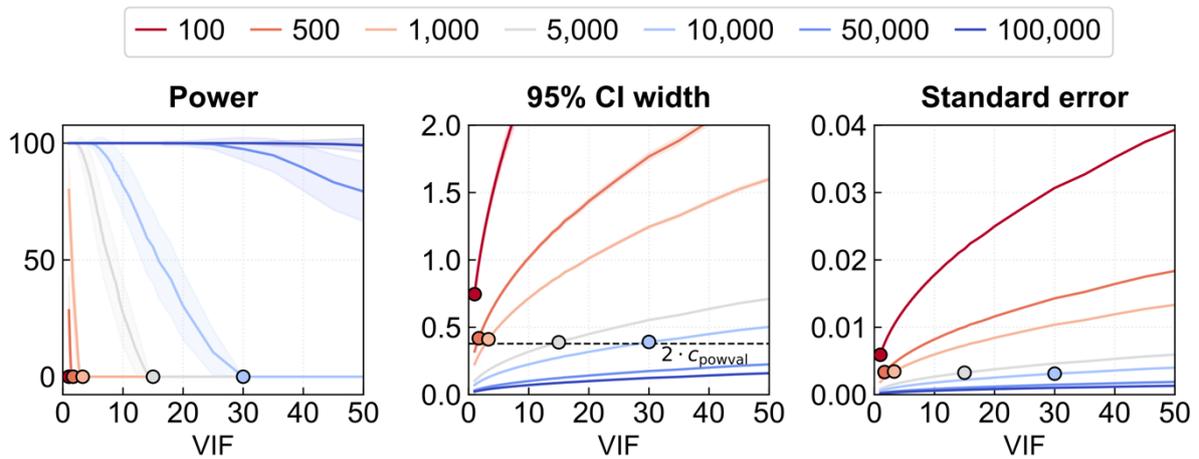

**Figure S4** | Mean power (in %), mean 95% confidence interval width, and mean standard error (SE) for $\hat{\beta}_{\text{main}}$ as a function of the collinearity, measured via the VIF, with all predictors being partially collinear. Each color corresponds to a different sample size, varying from 100 (deep red) to 100,000 (deep blue). Shaded areas represent the standard error around the mean estimates.

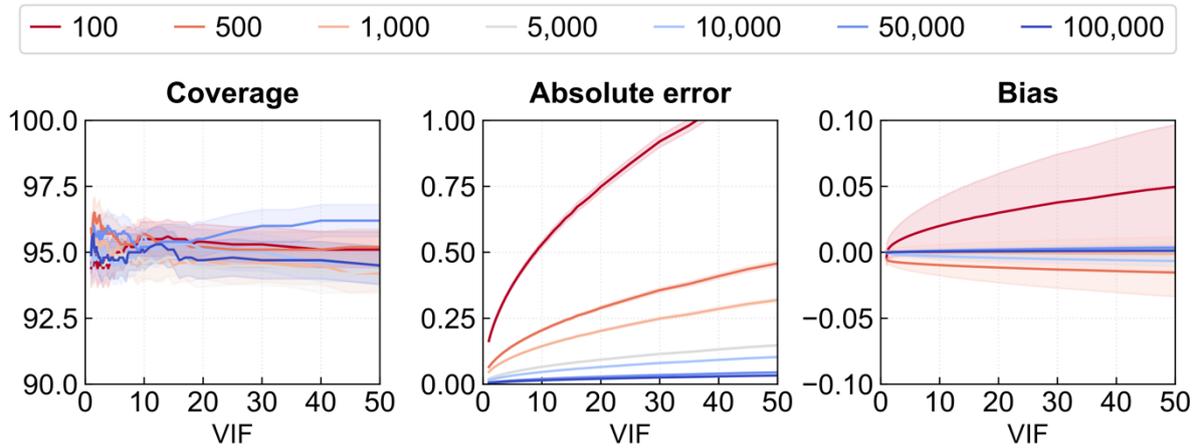

**Figure S5 |** Mean coverage (in %), mean absolute error, and mean bias for $\hat{\beta}_{\text{main}}$ as a function of the collinearity, measured via the VIF. Simulations were performed with 20 independent variables, all partially collinear. Each color represents a different sample size, ranging from 100 (deep red) to 100,000 (deep blue). Shaded areas represent the standard error around the mean estimates.

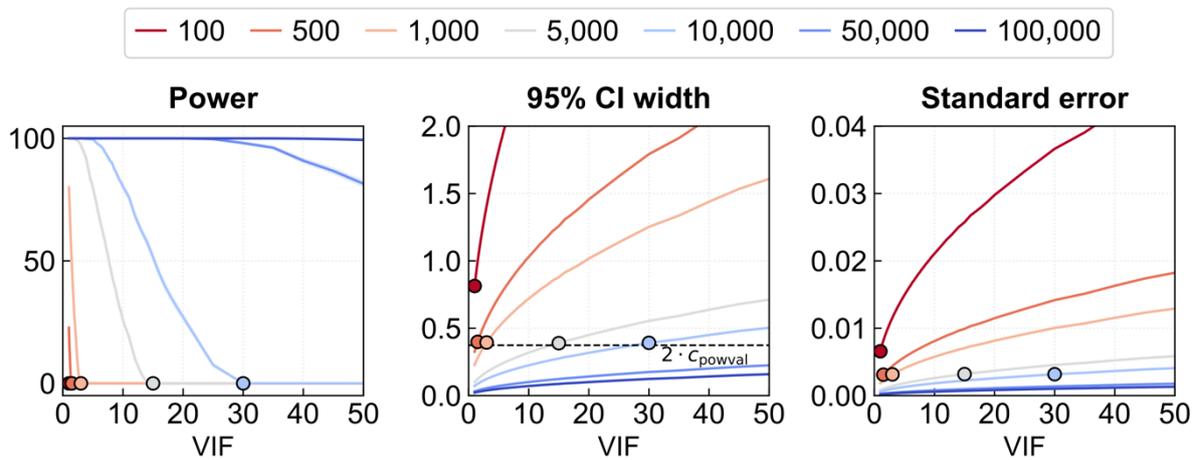

**Figure S6 |** Mean power (in %), mean 95% confidence interval width, and mean standard error (SE) for $\hat{\beta}_{\text{main}}$ as a function of the collinearity, measured via the VIF. Simulations were performed with 20 independent variables, all partially collinear. Each color represents a different sample size, ranging from 100 (deep red) to 100,000 (deep blue). Shaded areas represent the standard error around the mean estimates.

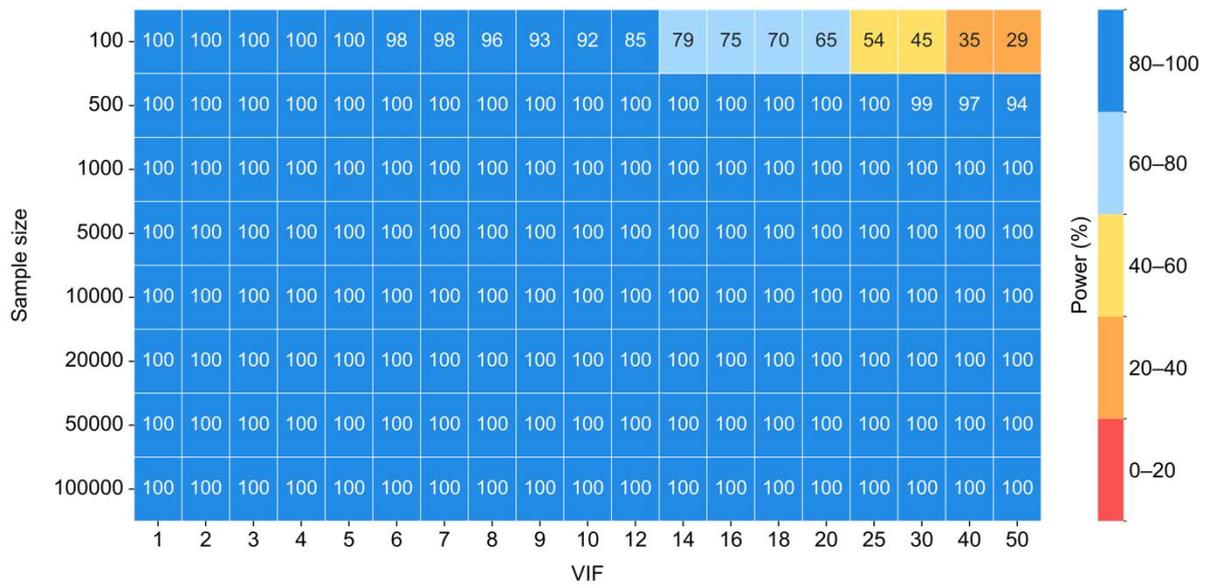

**Figure 7** | Mean traditional power (in %) for $\hat{\beta}_{\text{main}}$ as a function of the collinearity, measured via the VIF (x-axis), where $x_{\text{main}}$ and $x_1$ are partially collinear, and the sample size (y-axis), with $\boldsymbol{\beta_{\text{main}}} = \mathbf{2.0}$. Traditional power is defined as the proportion of simulations where the confidence interval does not include 0.

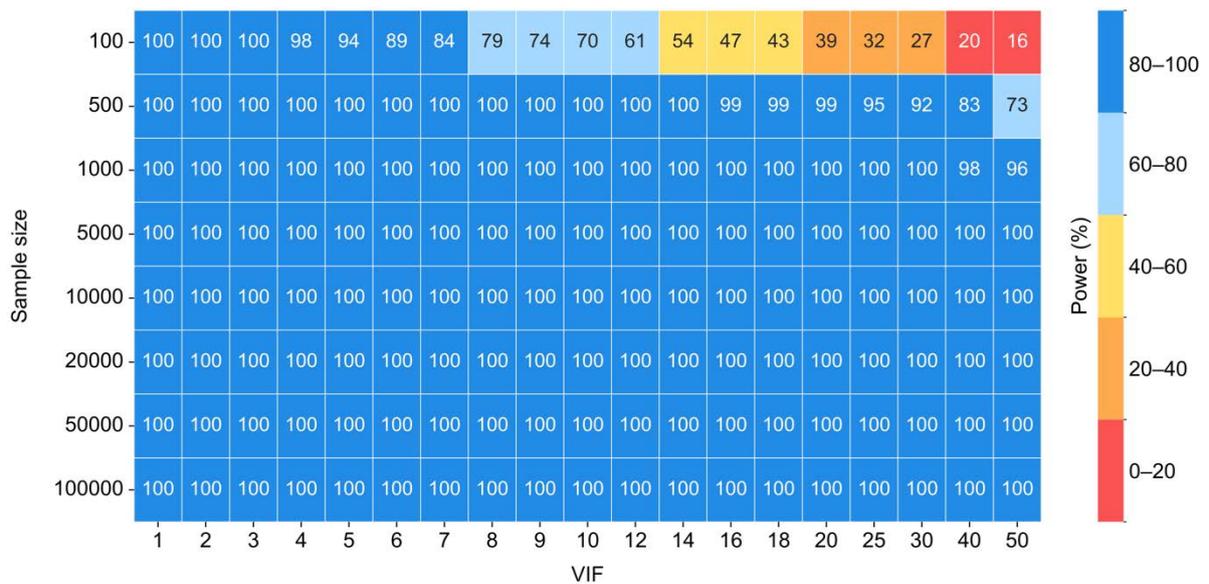

**Figure 8** | Mean traditional power (in %) for $\hat{\beta}_{\text{main}}$ as a function of the collinearity, measured via the VIF (x-axis), where $x_{\text{main}}$ and $x_1$ are partially collinear, and the sample size (y-axis), with $\boldsymbol{\beta_{\text{main}}} = \mathbf{1.5}$. Traditional power is defined as the proportion of simulations where the confidence interval does not include 0.

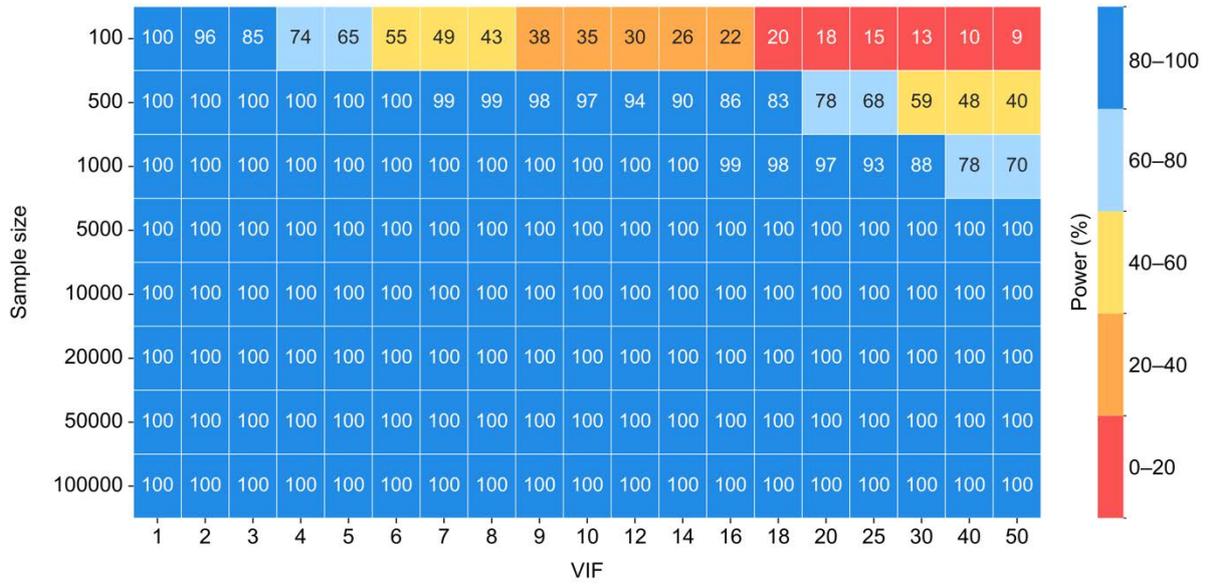

**Figure 9** | Mean traditional power (in %) for $\hat{\beta}_{\text{main}}$ as a function of the collinearity, measured via the VIF (x-axis), where $x_{\text{main}}$ and $x_1$ are partially collinear, and the sample size (y-axis), with $\boldsymbol{\beta_{\text{main}}} = \mathbf{1.0}$. Traditional power is defined as the proportion of simulations where the confidence interval does not include 0.

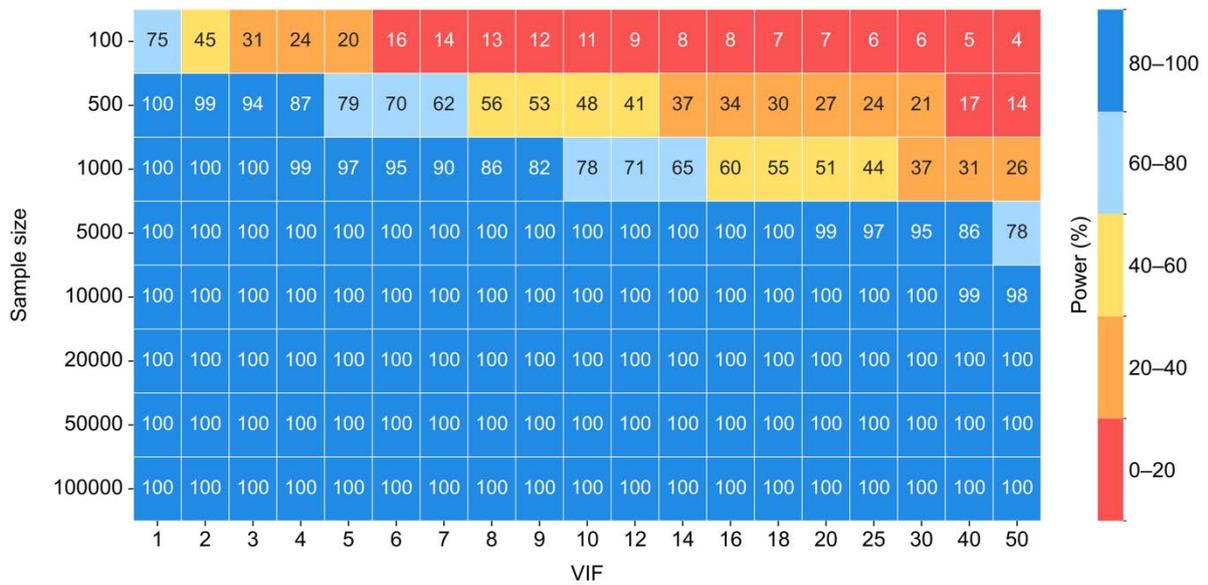

**Figure 10** | Mean traditional power (in %) for $\hat{\beta}_{\text{main}}$ as a function of the collinearity, measured via the VIF (x-axis), where $x_{\text{main}}$ and $x_1$ are partially collinear, and the sample size (y-axis), with $\boldsymbol{\beta_{\text{main}}} = \mathbf{0.5}$. Traditional power is defined as the proportion of simulations where the confidence interval does not include 0.

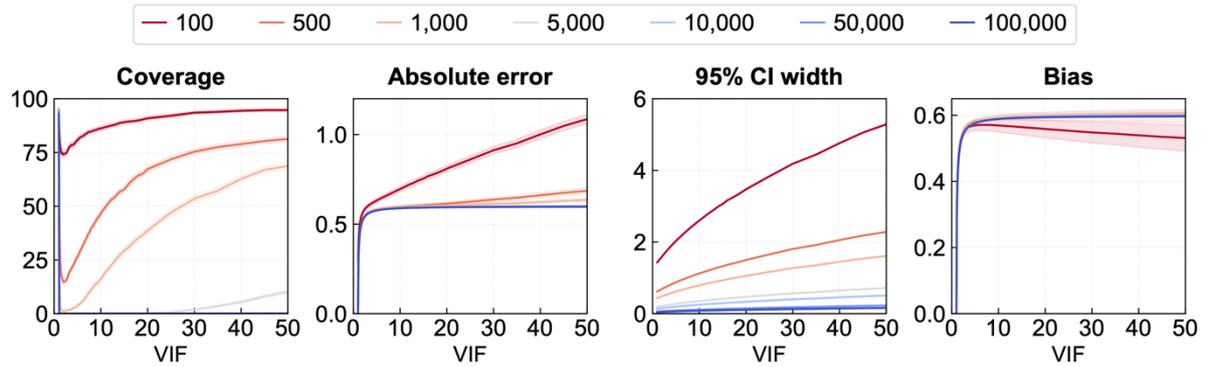

**Figure S11** | Mean coverage (in %), mean absolute error, mean 95% confidence interval width, and mean bias for $\hat{\beta}_{\text{main}}$ as a function of collinearity, measured via the VIF. The dataset is generated with all six predictors being partially collinear, after which one predictor ($x_4$) is omitted, introducing bias in the dataset. Each color represents a different sample size, ranging from 100 (deep red) to 100,000 (deep blue).

As shown in the figure, **coverage** initially decreases until approximately VIF=2 and then increases with higher collinearity. This trend occurs because the bias leads to a large **absolute error** (>0.5 around VIF>1.7) in $\hat{\beta}_{\text{main}}$, meaning that the estimated coefficient deviates considerably from the true value $\beta_{\text{main}} = 2$. While the **95 % CI width** also increases with collinearity, it does so more slowly than the absolute error, causing the true coefficient to fall outside the confidence interval more frequently at moderate VIF levels. Once the CI width expands sufficiently, the true coefficient is more often contained within the confidence intervals, leading to an improvement in coverage. Lower sample sizes exhibit higher coverage under collinearity because they produce wider confidence intervals, increasing the likelihood that the true coefficient falls within them. The **bias** rises rapidly to approximately 0.5 near VIF=2, and then converges to ~0.57. Notably, the bias is consistent across all sample sizes, except for N=100, which has a slightly lower bias. However, bias for N=100 converges to the same level as larger sample sizes when increasing the number of simulations per scenario, as also seen in Figure S1.

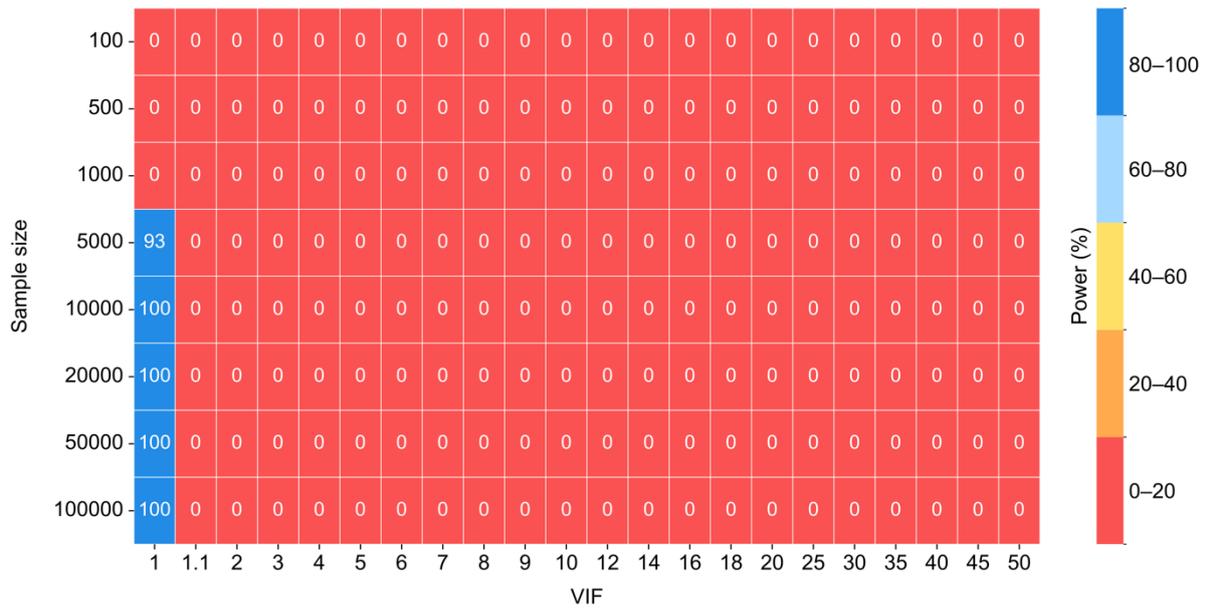

**Figure S12 |** Mean power (in %) for $\hat{\beta}_{\text{main}}$ as a function of collinearity, measured via the VIF (X-axis), and sample size (y-axis). The dataset is generated with all six predictors being partially collinear, after which one predictor ($x_4$) is omitted, introducing bias into the dataset.

The simulations only exhibit power at sample sizes of 5000 and higher, and only in the absence of collinearity. Once collinearity is introduced (VIF=1.1), power drops to zero immediately. This drop in power is due to two reasons. First, bias results in a widening of the confidence intervals (Figure S13), making it more likely that the 95% CI does not fall within $\beta_{\text{main}} \pm c_{\text{powval}}$. Second, the bias in the dataset leads to bias in the estimated coefficient $\hat{\beta}_{\text{main}}$ (Figure S11), shifting away the estimate from the $\beta_{\text{main}} \pm c_{\text{powval}}$ range. This effect is schematically illustrated in Figure S13.

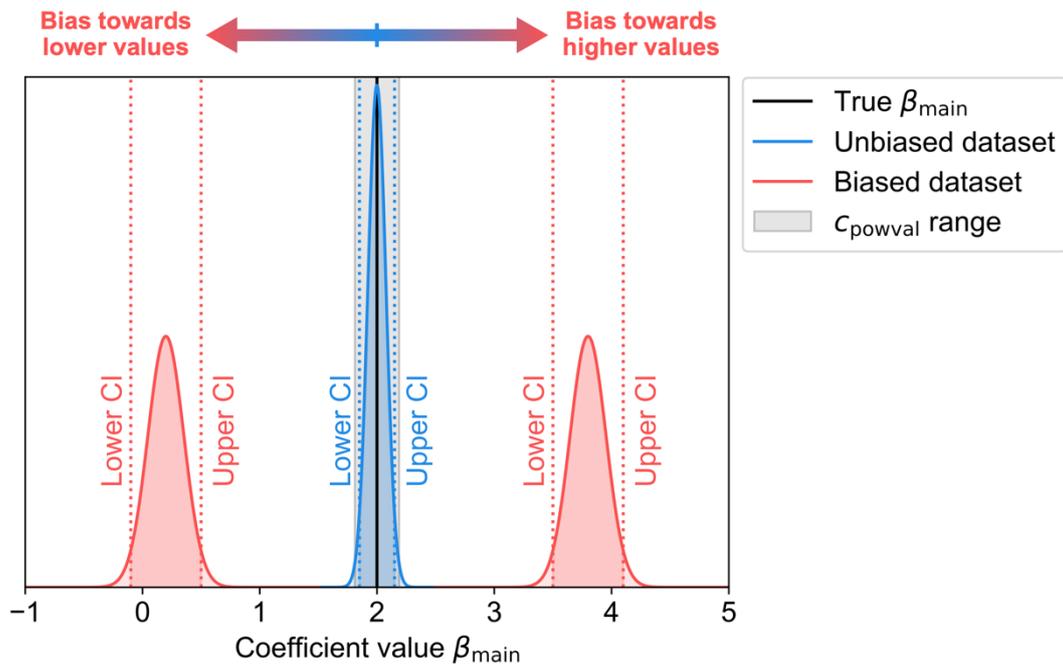

**Figure S13** | Schematic representation of why bias results in a loss of power. For an unbiased dataset, the 95% confidence interval (CI) of $\hat{\beta}_{main}$ (blue curve) falls within the $\beta_{main} \pm c_{powval}$ range, which is defined as having power. However, when bias is introduced, the estimates of $\hat{\beta}_{main}$ shift away from the true value and have wider CIs (red curves). Both effects contribute to the 95% CI of $\hat{\beta}_{main}$ falling outside the $\beta_{main} \pm c_{powval}$ range, resulting in a loss of power.

This figure also illustrates how bias can result to underestimation or overestimation of a true effect, potentially causing Type I and Type II errors.

**Table S1** | True coefficients used in the simulations with 20 independent variables.

| Coefficient | Value |
|---|---|
| intercept | 10.0 |
| $\beta_{\mathrm{main}}$ | 2.0 |
| $\beta_1$ | 1.3 |
| $\beta_2$ | 1.5 |
| $\beta_3$ | 6.0 |
| $\beta_4$ | 3.0 |
| $\beta_5$ | 1.0 |
| $\beta_6$ | 6.6 |
| $\beta_7$ | 0.7 |
| $\beta_8$ | 3.1 |
| $\beta_9$ | 2.6 |
| $\beta_{10}$ | 7.5 |
| $\beta_{11}$ | 6.9 |
| $\beta_{12}$ | 9.0 |
| $\beta_{13}$ | 1.3 |
| $\beta_{14}$ | 4.5 |
| $\beta_{15}$ | 0.8 |
| $\beta_{16}$ | 2.6 |
| $\beta_{17}$ | 5.3 |
| $\beta_{18}$ | 0.8 |
| $\beta_{19}$ | 2.4 |